\documentclass[conference]{IEEEtran}
\IEEEoverridecommandlockouts
\usepackage{cite}
\usepackage{amsmath,amssymb,amsfonts}
\usepackage{algorithmic}
\usepackage{graphicx}
\usepackage{textcomp}
\usepackage{xcolor}
\usepackage{fancyhdr}
\usepackage{titlesec}
\usepackage{multirow}
\usepackage{tabularx}
\usepackage{enumitem}
\usepackage{array}
\usepackage{listings}
\usepackage{caption}
\usepackage{subcaption}
\usepackage{setspace} 

\def\BibTeX{{\rm B\kern-.05em{\sc i\kern-.025em b}\kern-.08em
    T\kern-.1667em\lower.7ex\hbox{E}\kern-.125emX}}
\begin{document}

\title{Impact of White-Box Adversarial Attacks on Convolutional Neural Networks
\thanks{979-8-3503-5326-6/24/\$31.00 ©2024 IEEE}
}


\author{\IEEEauthorblockN{1\textsuperscript{st} Rakesh Podder}
\IEEEauthorblockA{\textit{Department of Computer Science} \\
\textit{Colorado State University}\\
Fort Collins, CO, USA, 80523 \\
rakesh.podder@colostate.edu}
\and
\IEEEauthorblockN{2\textsuperscript{nd} Sudipto Ghosh}
\IEEEauthorblockA{\textit{Department of Computer Science} \\
\textit{Colorado State University}\\
Fort Collins, CO, USA, 80523 \\
sudipto.ghosh@colostate.edu}
}




\maketitle

\begin{abstract}
Autonomous vehicle navigation and healthcare diagnostics are among the many fields where the reliability and security of machine learning models for image data are critical. We conduct a comprehensive investigation into the susceptibility of Convolutional Neural Networks (CNNs), which are widely used for image data, to white-box adversarial attacks. We investigate the effects of various sophisticated attacks---Fast Gradient Sign Method, Basic Iterative Method, Jacobian-based Saliency Map Attack, Carlini \& Wagner, Projected Gradient Descent, and DeepFool---on CNN performance metrics, (e.g., loss, accuracy), the differential efficacy of adversarial techniques in increasing error rates, the relationship between perceived image quality metrics (e.g., ERGAS, PSNR, SSIM, and SAM) and classification performance, and the comparative effectiveness of iterative versus single-step attacks. Using the MNIST, CIFAR-10, CIFAR-100, and Fashion\_MNIST datasets, we explore the effect of different attacks on the CNNs performance metrics by varying the hyperparameters of CNNs. Our study provides insights into the robustness of CNNs against adversarial threats, pinpoints vulnerabilities, and
underscores the urgent need for developing robust defense mechanisms to protect CNNs and ensuring their trustworthy deployment in real-world scenarios.


\end{abstract}


%
\IEEEpeerreviewmaketitle

\section{Introduction}
\label{introduction}

In the landscape of escalating cyber warfare, adversarial attacks on machine learning (ML) models have emerged as a sophisticated vector for undermining AI-driven systems. The inherent susceptibility of ML algorithms to specially crafted inputs that can lead to incorrect outputs, known as adversarial examples, has introduced a pressing challenge to the field of cybersecurity. The use of ML models in critical applications, such as autonomous vehicles~\cite{qayyum2020securing},
 healthcare diagnostics~\cite{kalaiselvi2020machine}, surveillance~\cite{cameron2019design}, and XR~\cite{vzidek2021cnn}, has become prevalent. The trust placed by end-users in various industry domains, healthcare, and governments on the reliability and security of AI-driven systems is fundamental to their widespread adoption.

Adversarial attacks have rapidly evolved from theoretical considerations to practical threats. These attacks leverage knowledge of the ML model's structure and data processing to introduce subtle perturbations, often imperceptible to humans but catastrophic for the model's decision-making accuracy. The consequences of successful adversarial attacks can range from trivial misclassifications to life-threatening situations~\cite{haffar2021explaining}. Therefore, understanding and mitigating these attacks are not just academic exercises; they are urgent requirements for the safe deployment of ML in real-world scenarios.

%
%

The goal of this paper is to conduct a systematic evaluation of various white-box adversarial attacks~\cite{ebrahimi2017hotflip}, where the attacker has complete visibility into the model's architecture, parameters, and training data, on generic neural network models for images. We use a curated set of images such as the MNIST~\cite{lecun2010mnist}, CIFAR-10, CIFAR-100~\cite{krizhevsky2009learning}, and Fashion\_MNIST~\cite{xiao2017fashion} datasets processed by a Convolutional Neural Network (CNN). The datasets take into account the variety and complexity required to challenge the CNNs under test. We identify the intrinsic vulnerabilities of CNNs when exposed to white-box attacks such as Fast Gradient Sign Method (FGSM)~\cite{goodfellow2014explaining}, Basic Iterative Method (BIM)~\cite{kurakin2018adversarial,alexey2016adversarial}, Jacobian-based Saliency Map Attack (JSMA)~\cite{papernot2016limitations}, Carlini \& Wagner (C\&W)~\cite{carlini2017towards}, Projected Gradient Descent (PGD)~\cite{madry2017towards}\cite{ayas2022projected}, and DeepFool~\cite{moosavi2016deepfool}.

Through a range of test scenarios that simulate attacks using the Adversarial Robustness Toolbox (ART) library~\cite{nicolae2018adversarial}, we discern how different performance metrics are affected, specifically focusing on the accuracy and loss incurred by the model under adversarial conditions. We quantify the degradation of performance in CNNs and investigate the robustness of these networks against such exploits. We also evaluate the impact on the image quality and integrity, with a specific focus on the assessment of widely recognized image analysis metrics such as ERGAS~\cite{du2007performance}, PSNR~\cite{alimuddin2012assessment}, SSIM~\cite{wang2004image}, and SAM~\cite{alparone2007comparison}. 


The rest of the paper is organized as follows. Section~\ref{background} provides a brief background on the attacks used in this paper. Section~\ref{gqm} outlines the evaluation goals, research questions, and metrics. Section~\ref{study-design} describes the study design and experimental environment. The results are presented in Section~\ref{results} and discussed in Section~\ref{discussion}. Section~\ref{related-work} summarizes related work. Section~\ref{conclusion} summarizes our conclusions and outlines directions for future work.

\section{Background}
\label{background}

Our study selected the following sophisticated white-box adversarial attacks based on their relevance to CNN models and image data,  prevalence in the current research literature, and real-world applicability~\cite{ebrahimi2017hotflip,wu2020making,roshan2024untargeted}. 

\begin{itemize}

    \item  \textbf{Fast Gradient Sign Method (FGSM)}: 
    New images that are classified incorrectly are created by leveraging the gradients of the loss with respect to the input image~\cite{goodfellow2014explaining}. Even though the method is straightforward, it is powerful in demonstrating the vulnerability of neural networks to slight, often imperceptible, changes in the input data.

    \item \textbf{Basic Iterative Method (BIM)}: BIM~\cite{kurakin2018adversarial,alexey2016adversarial} is an extension of FGSM. It iteratively applies the gradient sign attack with small steps, allowing for finer control over the perturbation process and often results in more effective adversarial examples.

    \item \textbf{Jacobian-based Saliency Map Attack (JSMA)}: JSMA~\cite{papernot2016limitations} uses the model's Jacobian matrix to determine which pixels in the input image to alter to change the classification outcome. The method is more refined than FGSM and BIM, and attempts to change the least number of pixels, thus making the alterations less detectable.

    \item \textbf{Carlini \& Wagner (C\&W)}: The C\&W attack~\cite{carlini2017towards} is an effective method that formulates adversarial example creation as an optimization problem. It aims to find the smallest perturbation that can mislead the CNN model, ensuring that the adversarial examples remain as close as possible to the original images. 

    \item \textbf{Projected Gradient Descent (PGD)}: PGD is a well-known variation of the BIM, distinguished by its initialization with uniform random noise~\cite{madry2017towards}. Similarly, the Iterative Least-likely Class Method (ILLC)~\cite{kurakin2018adversarial} bears a resemblance to BIM, with a key difference being its targeting of the least likely class to maximize the cross-entropy loss making it more effective than FGSM, JSMA and C\&W~\cite{ye2021thundernna}.

    \item \textbf{DeepFool}: This algorithm iteratively perturbs the input image in a way that is intended to cross the decision boundary of the classifier~\cite{moosavi2016deepfool}. It aims to be as efficient as possible, resulting in minimal perturbation~\cite{madry2017towards}.

\end{itemize}

\section{Evaluation Goals, Questions, and Metrics}
\label{gqm}



The main objective of this evaluation is to measure and understand the impact of white-box adversarial attacks on the performance and reliability of CNNs in the context of image processing. We aim to establish a rigorous testing method for detecting vulnerabilities within CNNs and to quantify the effectiveness of adversarial attacks in degrading model performance. The following  goals guide our evaluation:

\vspace{0.1in}
\noindent
\textbf{Goals:}

\begin{enumerate}
    \item Assess the impact of adversarial attacks on the accuracy and integrity of the image classification process.
    \item Identify the attack methodologies that result in the most significant degradation of performance metrics.
    \item Provide insights into the development of more robust CNN architectures and training processes.
\end{enumerate}

\vspace{0.1in}
\noindent
\textbf{Questions:}

\begin{enumerate}
    \item How do various white-box adversarial attacks affect the classification accuracy of CNNs?
    \item Which adversarial attack is most effective in inducing the highest error rates?
    \item What is the relationship between perceived image quality and classification performance of CNNs under attack?
    \item How does the iterative nature of certain attacks (e.g., BIM, PGD) compare to single-step attacks (e.g., FGSM) in terms of effectiveness?
\end{enumerate}

\vspace{0.1in}
\noindent
\textbf{Metrics:}

We use a combination of traditional CNN performance metrics and specialized image quality assessments:

\begin{itemize}

    \item \textbf{Loss}: This is a measure of how well the model performs from an error perspective. Specifically, it represents the \emph{cost} incurred for inaccurate predictions. In the code, the loss is calculated using \texttt{sparse\_categorical\_crossentropy}, which is a common loss function for classification tasks. It compares the predicted probability distribution (output of the \texttt{softmax} function in the last layer) with the true distribution, where the true distribution is the label of the class that the input image belongs to. A lower loss indicates better performance of the model, as it means the model's predictions are closer to the true labels.
    \item \textbf{Accuracy}: This is a measure of the proportion of correctly predicted instances out of all predictions made. In a classification task like MNIST (which involves classifying images of handwritten digits into 10 classes, from 0 to 9), the accuracy is calculated by the number of images correctly classified divided by the total number of images classified. Higher accuracy means that the model has better predictive performance.
    \item \textbf{Relative Dimensionless Global Error in Synthesis (ERGAS):} This is a global measure of image fidelity, with lower values indicating better synthesis quality~\cite{du2007performance}.
    \begin{align*}
                \text{ERGAS} = 100 \cdot \sqrt{\frac{1}{d} \sum_{i=1}^{N} \left( \frac{\text{RMSE}_i}{\mu_i} \right)^2}
    \end{align*}
    \begin{itemize}
    \item $d$ is the scale factor between the spatial resolutions of the original and the processed image (often set to 1 for images of the same resolution).
    \item $N$ is the number of bands.
    \item $\text{RMSE}_i$ is the Root Mean Square Error of the $i$th band.
    \item $\mu_i$ is the mean of the $i$th band of the original image.
\end{itemize}

    \item \textbf{Peak Signal-to-Noise Ratio (PSNR):} This is a measure of peak error, with higher values indicating smaller differences between original and perturbed images. PSNR is calculated using the maximum pixel value ($L$) and the Mean Squared Error ($MSE$) between the original ($I$) and corrupted ($K$) images. $M$ and $N$ are the number of rows and columns respectively in the images~\cite{alimuddin2012assessment}.
    \begin{align*}
        \text{PSNR} = 20 \cdot \log_{10}(L) - 10 \cdot \log_{10}(\text{MSE})\\
        \text{MSE} = \frac{1}{MN} \sum_{i=1}^{M} \sum_{j=1}^{N} \left( I(i,j) - K(i,j) \right)^2
    \end{align*}
    \item \textbf{Structural Similarity Index (SSIM):} This is a perception-based model that considers changes in texture, contrast, and luminance~\cite{wang2004image}.
    \begin{align*}
        \text{SSIM}(x, y) = \frac{(2\mu_x\mu_y + c_1)(2\sigma_{xy} + c_2)}{(\mu_x^2 + \mu_y^2 + c_1)(\sigma_x^2 + \sigma_y^2 + c_2)}
    \end{align*}
    \begin{itemize}
    \item $x, y$ are the windowed images being compared.
    \item $\mu_x, \mu_y$ are the averages of $x$ and $y$.
    \item $\sigma_x^2, \sigma_y^2$ are the variances of $x$ and $y$.
    \item $\sigma_{xy}$ is the covariance of $x$ and $y$.
    \item $c_1, c_2$ are variables to stabilize division with a weak denominator.
    \end{itemize}
    \item \textbf{Spectral Angle Mapper (SAM):} It is a measure of the spectral similarity between two images, with lower values indicating higher similarity~\cite{alparone2007comparison}. It measures the angle between the spectral vectors $\mathbf{a}$ and $\mathbf{b}$.
    
    \begin{align*}
        \text{SAM} = \cos^{-1} \left( \frac{\mathbf{a} \cdot \mathbf{b}}{\|\mathbf{a}\| \|\mathbf{b}\|} \right) \\
    \end{align*}
    
\end{itemize}

The above metrics are calculated before and after the application of each adversarial attack.

\begin{itemize}
  \item \textbf{Pre-attack Performance:} We establish the baseline values of loss, accuracy, ERGAS, PSNR, SSIM, and SAM.
  \item \textbf{Post-attack Performance:} The same metrics are re-assessed post-adversarial attack to evaluate the impact.
  \item \textbf{Adversarial Success Rate:} We record the rate at which adversarial inputs successfully deceive the CNN.
  \item \textbf{Robustness Threshold:} We identify the minimal perturbation magnitude necessary to compromise the model.
\end{itemize}


\section{Study Design}
\label{study-design}

Our study incorporated the following steps, which we illustrate using the FGSM attack for lack of space. We developed a python script using the \textit{TensorFlow} and \textit{Adversarial Robustness Toolbox (ART)} libraries to (1)~create \& train a neural network on the MINST~\cite{lecun2010mnist}, CIFAR-10, CIFAR-100~\cite{krizhevsky2009learning}, \& Fashion\_MNIST~\cite{xiao2017fashion} datasets,  (2)~generate adversarial examples, 
and (3)~evaluate the models' performance on the adversarial examples. 
We loaded and preprocessed (normalized) the image datasets such as MNIST, CIFAR-10, CIFAR-100, and Fashion\_MNIST. 

For \textbf{model selection and preparation}, we used the TensorFlow API to create custom CNN models with different hyperparameter values (e.g., \textit{number of neurons}, \textit{dropout rate}, \textit{number of classes}, and \textit{optimizers}). For example, for the FGSM attack, we created a simple neural network model using \textit{TensorFlow's Keras API}. The model consists of a Flatten layer that converts each 28x28 MNIST type images into a 784 element vector, followed by a Dense layer with 128 nodes, a Dropout layer that randomly sets 20\% of the input units to 0 during training, and a final Dense layer with 10 nodes corresponding to the 10 possible digits (0-9). For CIFAR-10 \& CIFAR-100, the TensorFlow model is a CNN for 32x32 pixel RGB images, featuring three convolutional layers with ReLU activations for feature extraction—first with 32 filters, followed by two layers with 64 filters each, interspersed with 2x2 max pooling for dimensionality reduction. After the convolutional layers, it employs a flattening step, a dense layer of 64 units (ReLU activation), and concludes with a 10-unit softmax output layer for classifying into 10 categories.

We compiled the model using the \textit{adam} optimizer and the \textit{sparse\_categorical\_crossentropy} loss function, and then trained on the training images and labels for 5 epochs. We evaluated the trained model on the test images and labels to get the baseline loss and accuracy. Training for 5 epochs is sufficient to achieve a reasonable balance between training time and performance, allowing the model to learn effectively without overfitting. Additionally, running the model for more epochs could lead to only marginal improvements in performance metrics, as the model typically converges within the first few epochs.

After defining and training the model, we wrapped it within an ART classifier, such as \textit{TensorFlowV2Classifier} for TensorFlow models, specifying necessary hyperparameters like the number of classes, input shape, and loss object. For each attack type, we created an instance of the corresponding ART attack class, configuring it with relevant parameters (e.g., \textit{eps} for FGSM, \textit{max\_iter} for PGD). For example, for the input to the FGSM attack, we created ART's \textit{TensorFlowV2Classifier} using the trained model. ART's \textit{FastGradientMethod} attack is created using the classifier and an epsilon value of 0.1.

We performed \textbf{adversarial example generation} by leveraging state-of-the-art techniques to mislead the CNNs while preserving image quality. We implemented a \textit{generate} method that takes the attack instances and passes the original inputs. This process, although slightly varied in parameters and attack initialization, follows the same basic steps across different adversarial techniques, enabling the evaluation of model robustness under various types of adversarial conditions.

The \textbf{evaluation method} calculated the loss and accuracy on both the original and adversarial examples through the model's \textit{evaluate} method. This method computes the loss and accuracy metrics by comparing the model's predictions on the input images against the true labels. This process involves feeding the perturbed images into the model and calculating the metrics to assess how well the model performs on these adversarial inputs. A large decrease in accuracy or an increase in loss indicates that the adversarial attack was successful in degrading the model's performance.

\textbf{Statistical analysis} of the adversarial attack's impact was performed by comparing various metrics between the original and adversarial images. Using the \textit{sewar}~\cite{sewar} library, metrics such as ERGAS, PSNR, SSIM, and SAM were computed. 
The analysis was encapsulated in a DataFrame, providing a structured view of the impact across different metrics, thereby facilitating an understanding of the adversarial attack's effectiveness in degrading image quality and model performance.

\begin{equation*}
\begin{small}
\text{EF} = 
\left[
    \begin{array}{c}
        \text{loss} \\
        \text{accuracy} \\
        \text{ERGAS} \\
        \text{PSNR} \\
        \text{SSIM} \\
        \text{SAM}
    \end{array}
\right]
\cdot
\left( 
    \left[
        \begin{array}{c}
            \text{FGSM} \\
            \text{JSMA} \\
            \text{C\&W} \\
            \text{PGD} \\
            \text{DeepFool} \\
            \text{BIM}
        \end{array}
    \right]
    \times
    \left[
        \begin{array}{c}
            \text{MNIST} \\
            \text{CIFAR-10} \\
            \text{CIFAR-100} \\
            \text{Fashion\_MNIST}
        \end{array}
    \right]
\right)
\end{small}
\end{equation*}

The evaluation framework (EF) enables a comprehensive analysis by evaluating how each attack affects model performance across different types of data, quantified through metrics for aspects such as accuracy, error, and image quality. 


We used Input Space Partitioning (ISP) and Base Choice Coverage (BCC)~\cite{ammann2008introduction} in our test method to systematically explore the model's vulnerability across various configurations and adversarial scenarios. ISP facilitates a detailed examination of the model's input space, partitioning it down into multiple blocks for a nuanced analysis of vulnerability using various characteristics. Partitioning allows us to uncover a wider range of weaknesses by examining how different types of inputs can influence the model. BCC extends this analysis by first identifying the base choice for each characteristic that was used to partition the domain of an input variable, and then creating combinations of the input partitions, starting with all the base choices and then by varying one choice at a time.

\section{Results}
\label{results}
We ran our evaluation on a 3.1 GHz
Dual-Core Intel Core i5 processor, with 8 GB 2133 MHz,
and LPDDR3 memory. 

\subsection{Performance and Image Quality Metrics}

\begin{figure*}[!htb]
     \centering
     \begin{subfigure}[b]{0.4\linewidth}
         \centering
         \includegraphics[width=\linewidth]{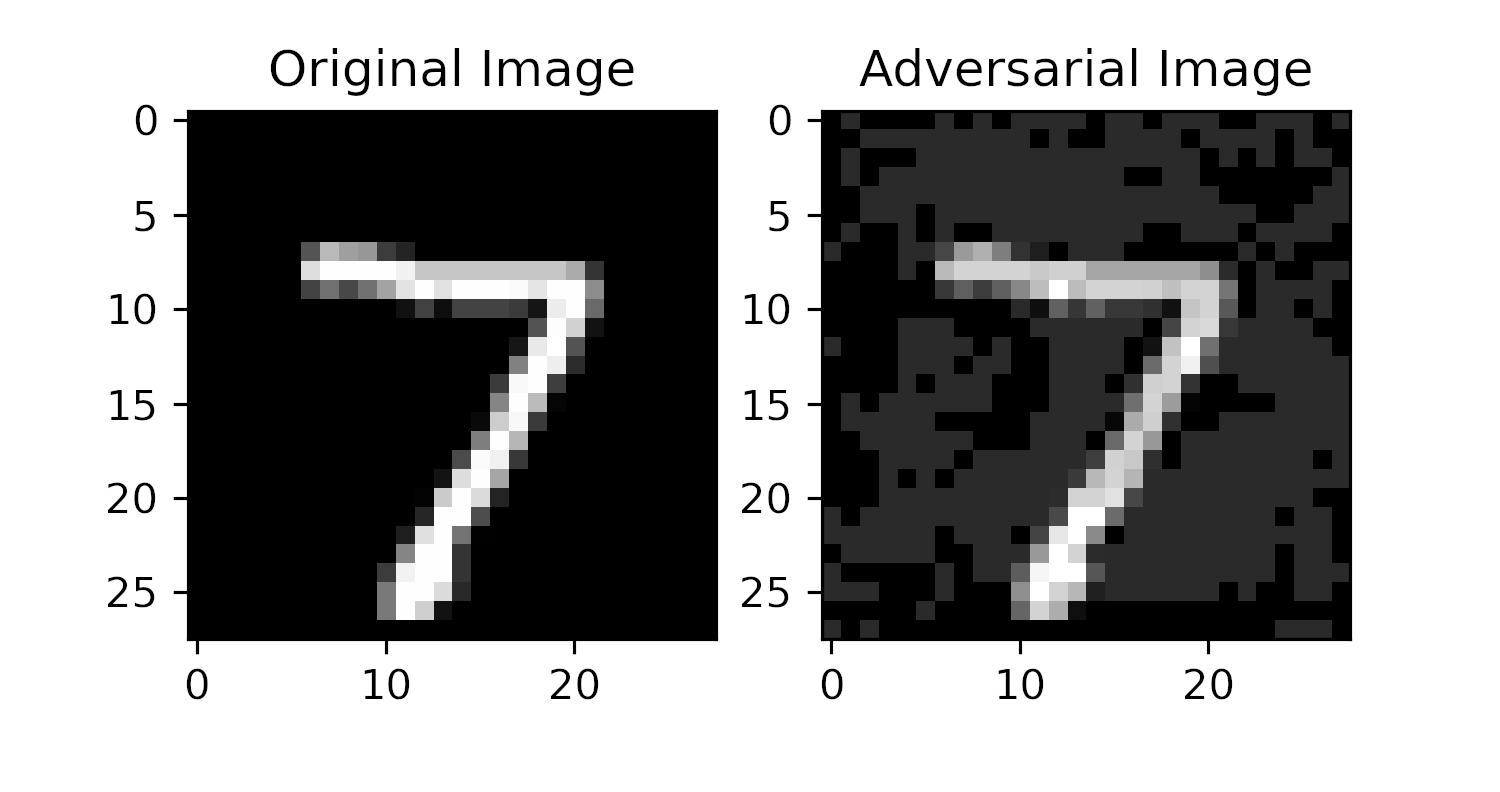}
         \caption{On MNIST}
         \label{fgsm-mnist}
     \end{subfigure}
     \hfill
     \begin{subfigure}[b]{0.4\linewidth}
         \centering
         \includegraphics[width=\linewidth]{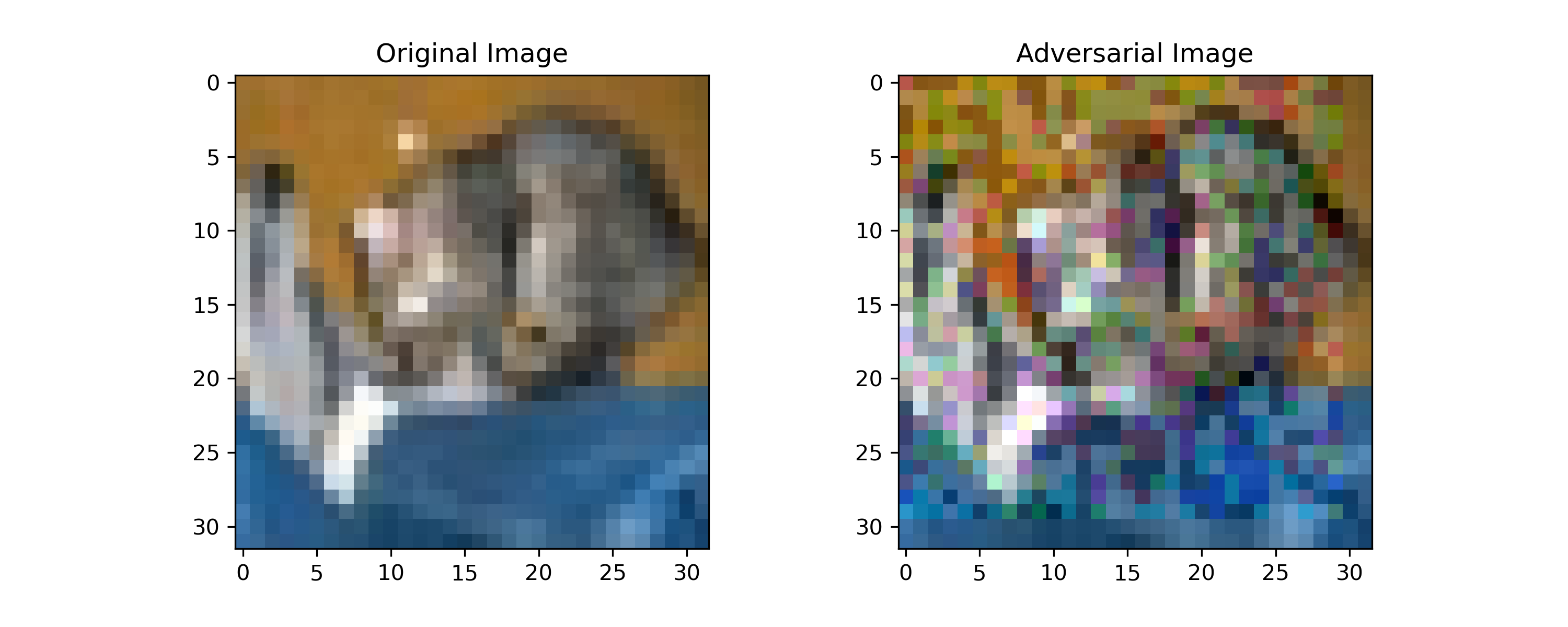}
         \caption{On CIFAR-10}
         \label{fgsm-cifar-10}
     \end{subfigure}
     \hfill
     \begin{subfigure}[b]{0.4\linewidth}
         \centering
         \includegraphics[width=\linewidth]{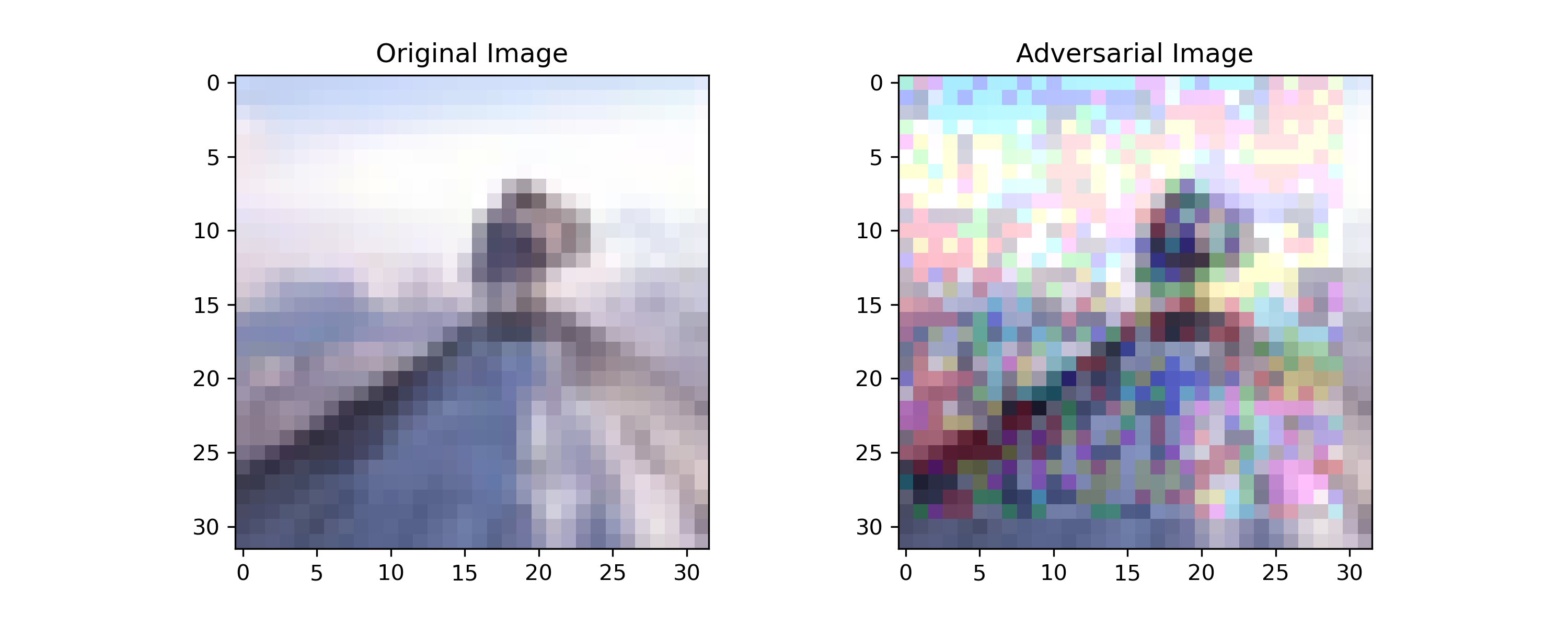}
         \caption{On CIFAR-100}
         \label{fgsm-cifar-100}
     \end{subfigure}
     \hfill
     \begin{subfigure}[b]{0.4\linewidth}
         \centering
         \includegraphics[width=\linewidth]{cifar100_comparison_image.png}
         \caption{On Fashion\_MNIST}
         \label{fgsm-fashion-mnist}
     \end{subfigure}
        \caption{Original and Compromised Images Generated from FGSM Attacks}
        \label{fig:fgsm-attacks}
\end{figure*}

For the FGSM attack described in Section~\ref{study-design}, the script trains a simple neural network on the various datasets, generates adversarial examples using the FGSM attack, evaluates the model's performance on the adversarial examples as shown in Table~\ref{tab:fsgm_datasets}, and uses \textit{matplotlib} to display an original and adversarial image side-by-side as shown in Figure~\ref{fig:fgsm-attacks}. 

\begin{table}[!htb]
\centering
\caption{Metrics From FGSM Adversarial Attacks on MNIST, CIFAR-10, CIFAR-100, \& Fashion\_MNIST}
\label{tab:fsgm_datasets}
\begin{tabular}{lllll}
\hline
\textbf{Metric} & \textbf{MNIST} &\textbf{CIFAR-10} & \textbf{CIFAR-100}&\textbf{Fashion}\\
& & & & \textbf{\_MNIST}\\\hline
Accuracy & 0.10 &0.12 & 0.04 & 0.19\\
loss & 3.08 & 6.35 & 6.96 & 6.48\\
ERGAS & 27.08 & 88.32 & 52.68 & 14.94\\
PSNR & 22.27  &18.60 & 8.99 & 4.28\\
SSIM & (0.882,  &0.71 & 0.38 & (0.123,\\
     &  0.945)  &     &      &  0.114)\\
SAM & 0.28 & 0.25 & 0.49 & 1.08\\
\hline
\end{tabular}
\end{table}

\begin{table*}[!htb]
\centering
\caption{Metric Evaluation Table for all Adversarial Attacks}
\label{tab:metric}
\begin{tabular}{lllllll}
\hline
\textbf{Metric} & \textbf{FGSM}&\textbf{DF}&\textbf{C\&W}&\textbf{PGD}&\textbf{JSMA}&\textbf{BIM}\\
\hline
Accuracy & 0.10& 0.00977& 0.977& 0.0135& 0.075&0.02\\
loss & 3.08& 2782.88& 0.074& 80.89& 0.975&16.5\\
ERGAS & 27.08&3254.48 &78.336 &79.73 &29.72 &26.56\\
PSNR & 22.5&2.617 &13.407 &13.52 &23.56 &22.55\\
SSIM & (0.89, 0.94)&(-0.13, -0.24)&(0.632, 0.71)&(0.62, 0.71)&(0.93, 0.934)&(0.896, 0.95)\\
SAM & 0.28&1.365663  &0.748 &0.749 &0.236 &0.27\\
\hline
\end{tabular}
\end{table*}

\begin{figure*}[!htb]
    \centering
    \begin{subfigure}[b]{0.4\linewidth}
        \includegraphics[width=\linewidth]{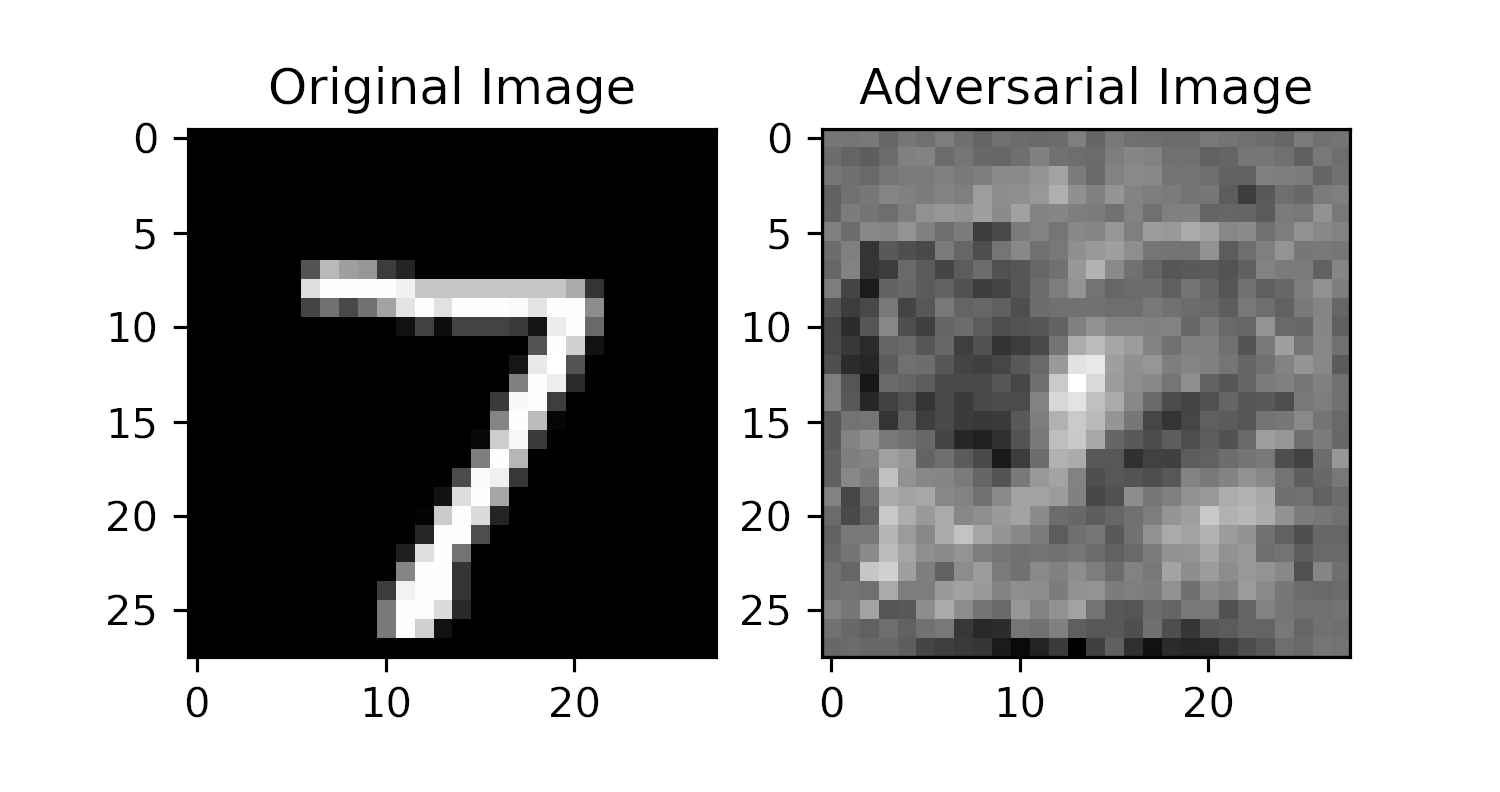}
        \caption{DeepFool attack}
    \end{subfigure}
    \hfill
    \begin{subfigure}[b]{0.4\linewidth}
        \includegraphics[width=\linewidth]{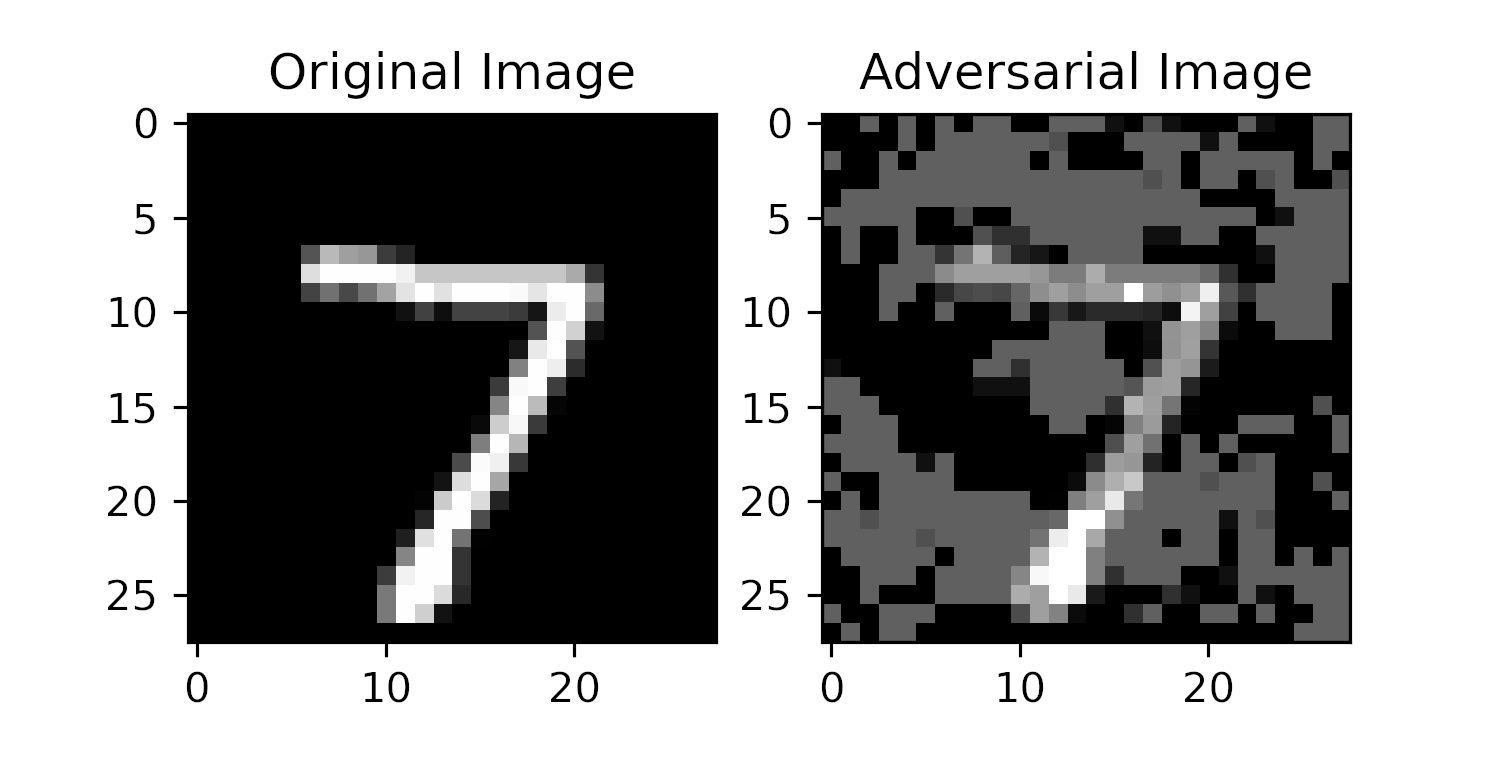}
        \caption{PGD attack}
    \end{subfigure}
    \hfill
    \begin{subfigure}[b]{0.4\linewidth}
        \includegraphics[width=\linewidth]{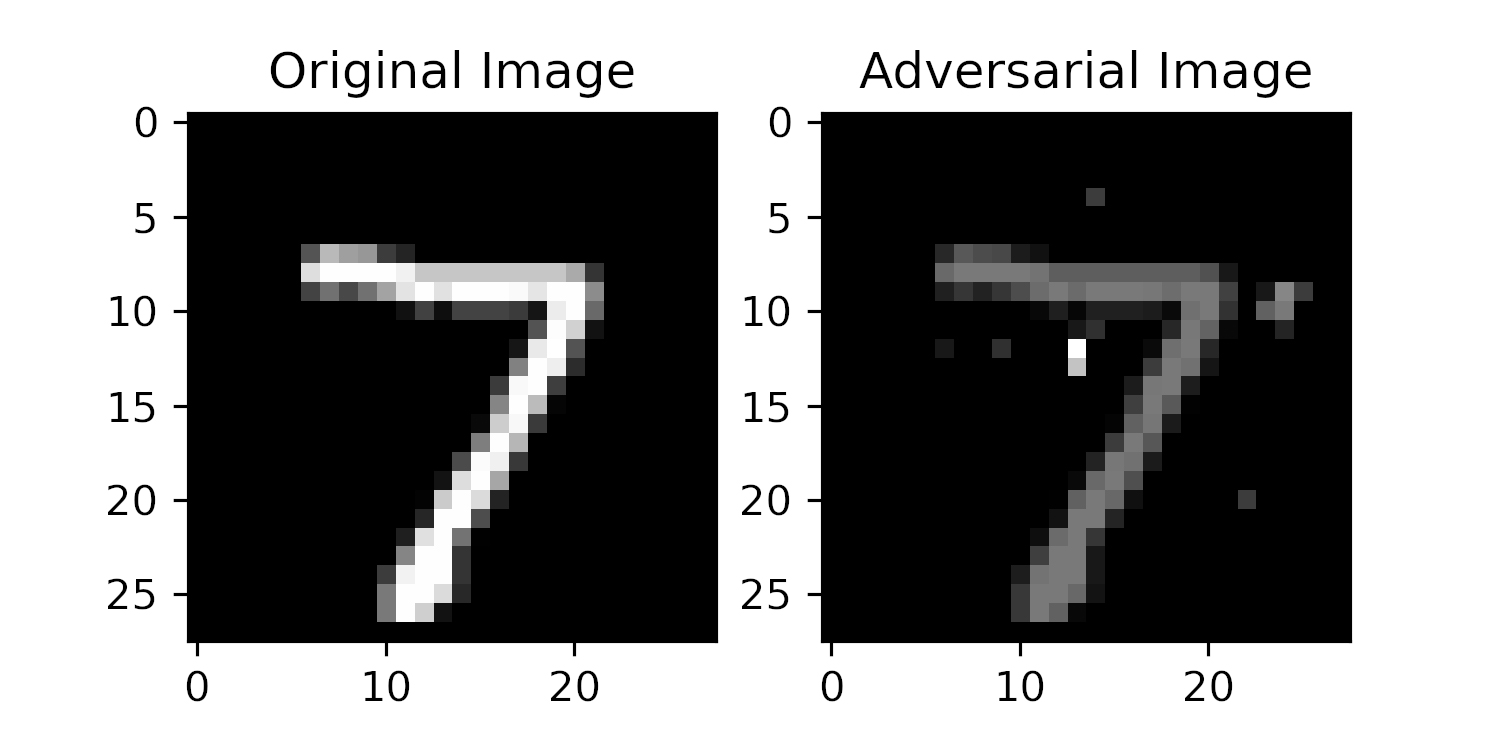}
        \caption{JSMA attack}
    \end{subfigure}
    \hfill
    \begin{subfigure}[b]{0.4\linewidth}
        \includegraphics[width=\linewidth]{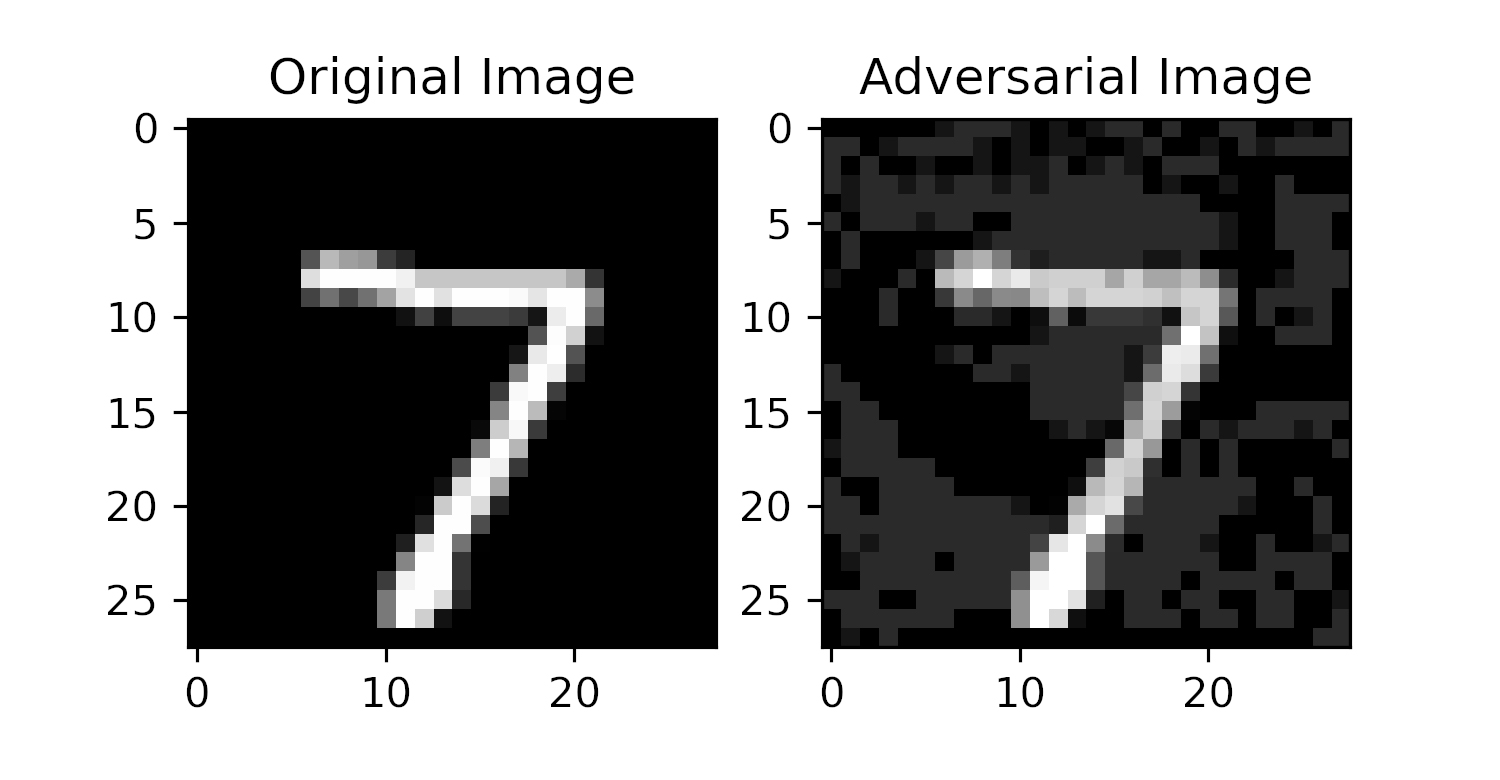}
        \caption{BIM attack}
    \end{subfigure}
    \caption{Original MNIST and Compromised Images Generated Using DeepFool, PGD, JSMA, and BIM.}
    \label{fig:attacks-on-cnn}
\end{figure*}

Table~\ref{tab:metric} details the outcomes of applying several adversarial techniques on the CNN metrics. Figure~\ref{fig:attacks-on-cnn} illustrates the practical effects of adversarial manipulations by presenting side-by-side comparisons of original and compromised images using the DeepFool, PGD, JSMA, and BIM attacks.

\begin{figure}[h!]
    \centering
    \includegraphics[width=\linewidth]{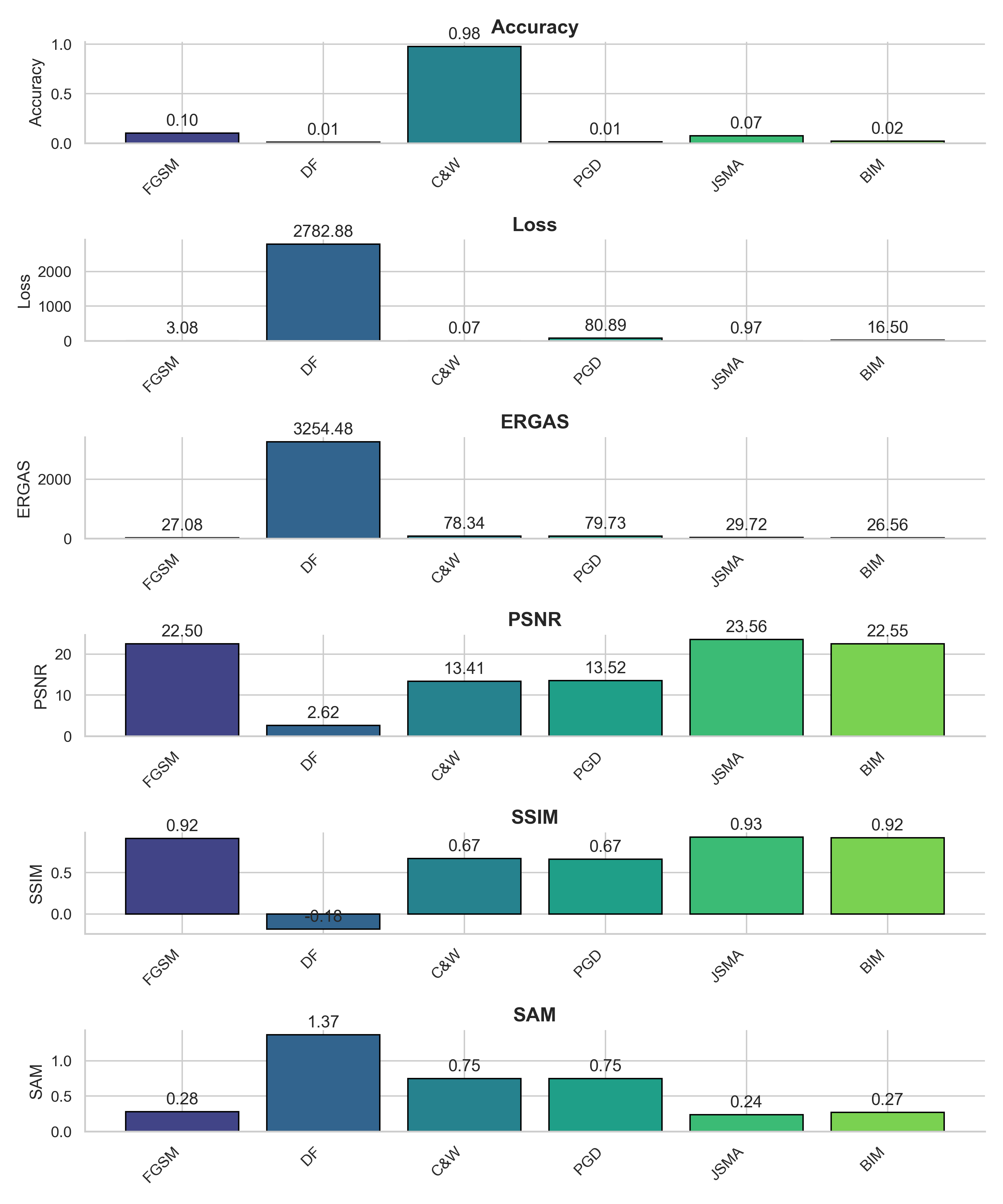}
    \caption{Effects of Adversarial Attacks on CNN}
    \label{fig:enter-label}
\end{figure}

\subsection{Input Space Partitioning and Base Choice Coverage}
The input variables analyzed using ISP are:
\begin{itemize}
    \item Number of Neurons: Numeric value.
    \item Dropout Rate: Numeric percentage.
    \item NB\_Classes: Numeric value.
    \item Optimizer: Alphanumeric value.
    \item Dataset Type: Image type data.

\end{itemize}
Table \ref{tab:isp} displays the results of ISP analysis in terms of the variables, the characteristics chosen to partition the input domains, the blocks in each partition, and representative values. For each input variable, the base choice is the block numbered 1 (e.g., a1, b1, c1, d1, and e1) as shown in Table~\ref{tab:bcc}.
BCC leverages these partitions to construct test cases, each designed to probe different combinations of input conditions. These parameters are crucial in a CNN because they directly influence the model's capacity to learn, generalize, and maintain robustness against adversarial attacks by affecting network complexity, regularization, classification capability, and optimization efficiency.

\begin{table*}[!htb]
\centering   
    \caption{Input Space Partitioning on the Hyperparameters}
    \label{tab:isp}
    \begin{tabular}{c|c|l|l}\hline
          Variables&  Characteristics& Partitions &Values\\
         \hline
          Number of Neurons ($N$)& Numeric: $0 < N < finite value$& a1: true&$N$ = 128\\
          & & a2: false&$N$ = NULL\\
  Dropout rate ($R$)& Numeric:
$0 <R < = 1$& b1: true&$R$ = 0.2\\
  & & b2: false&$R$ = - 0.3\\
  NB\_classes
($nb$)& Numeric: $0 < N < finite value$& c1: true&$nb \geq 2$\\
  & & c2: false&$nb < 2$\\
  Optimizer
($O$)& Alphanumeric& d1: true&$O$ = ``Adadelta
"\\
  & & d2: false&$O$ = NULL (None)\\
 Dataset type ($val$)& Image& e1: nonEmpty&$val$ = ``MNIST''\\
 & & e2: Empty&$val$ = NULL/Non-Image type\\\hline
 \end{tabular}
\end{table*}

\begin{table}[!htb]
    \centering
    \caption{Base Choice Coverage Table}
    \label{tab:bcc}
    \begin{tabular}{c|cllll}\hline
         Test&  Block 1 & Block 2 &Block 3 &Block 4 &Block 5 \\\hline
         $T_1$ (base)&  
     a1& b1& c1& d1& e1\\
 $T_2$& a2& b1& c1& d1& e1\\
 $T_3$& a1& b2& c1& d1& e1\\
 $T_4$& a1& b1& c2& d1& e1\\
 $T_5$& a1& b1& c1& d2& e1\\
 $T_6$& a1& b1& c1& d1& e2\\
 $T_7$& a2& b2& c1& d1& e1\\
 $T_8$& a2& b1& c2& d1& e2\\
 $T_9$& a1& b2& c1& d2& e2\\
 $T_{10}$& a1& b2& c2& d2& e1\\\hline
 \end{tabular}

\end{table}

\subsection{Testing}
\label{sec:08-test}

Tables~\ref{tab:t1} to \ref{tab:t4} present the results of evaluating the resilience of Convolutional Neural Networks (CNNs) to the FGSM attack under various conditions. Each table represents a distinct experiment focusing on altering one specific model parameter--number of neurons, dropout rate, number of classes, and optimizer--while keeping the others constant to observe its impact on the model's loss and accuracy.

\begin{table*}[!htb]
    \centering
        \caption{Testing the ISP on FGSM varying Number of Neurons (N).}
    \label{tab:t1}
    \begin{tabular}{c|cllll|ll}\hline
    Test Case&  N & R & nb & O & val &loss &accuracy\\\hline
         $TC_1$& 
     a1 = 128& b1 = 0.2& c2 = 10& d1 = Adadelta& e1 = MNIST& 3.08&0.105\\
 $TC_2$& a1 = 100& b1 = 0.2& c2 = 10& d1 = Adadelta& e1 = MNIST& 3.10&0.09\\
 $TC_3$& a1= 150& b1 = 0.2& c2 = 10& d1 = Adadelta& e1 = MNIST& 3.04&0.108\\
 $TC_4$& a1 = 500& b1 = 0.2& c2 = 10& d1 = Adadelta& e1 = MNIST& 2.9& 0.12\\
 $TC_5$& a1=1000& b1 = 0.2& c2 = 10& d1 = Adadelta& e1 = MNIST& 2.5&0.13\\\hline
 \end{tabular}
\end{table*}

\begin{table*}[!htb]
    \centering
        \caption{Testing the ISP on FGSM varying Dropout Rate (R).}
    \label{tab:t2}
    \begin{tabular}{c|cllll|ll}\hline
    Test Case&  N & R & nb & O & val &loss &accuracy\\\hline
         $TC_1$& 
     a1 = 128& b1 = 0.2& c2 = 10& d1 = Adadelta& e1 = MNIST& 3.08&0.105\\
 $TC_2$& a1 = 128& b1 = 0.02& c2 = 10& d1 = Adadelta& e1 = MNIST& 3.19&0.099\\
 $TC_3$& a1 = 128& b1 = 0.001& c2 = 10& d1 = Adadelta& e1 = MNIST& 3.8&0.06\\
 $TC_4$& a1 = 128& b1 = 0.5& c2 = 10& d1 = Adadelta& e1 = MNIST& 2.88& 0.11\\
 $TC_5$& a1 = 128& b1 = 0.2& c2 = 10& d1 = Adadelta& e1 = MNIST& 2.75&1.24\\\hline
 \end{tabular}

\end{table*}

\begin{table*}[!htb]
    \centering
        \caption{Testing the ISP on FGSM varying Number of Classes (nb). }
    \label{tab:t3}
    \begin{tabular}{c|cllll|ll}\hline
    Test Case&  N & R & nb & O & val &loss &accuracy\\\hline
         $TC_1$& 
     a1 = 128& b1 = 0.2& c2 = 10& d1 = Adadelta& e1 = MNIST& 3.08&0.105\\
 $TC_2$& a1 = 128& b1 = 0.2& c2 = 2& d1 = Adadelta& e1 = MNIST& 3.10&0.09\\
 $TC_3$& a1 = 128& b1 = 0.2& c2 = 50& d1 = Adadelta& e1 = MNIST& 3.04&0.108\\
 $TC_4$& a1 = 128& b1 = 0.2& c2 = 100& d1 = Adadelta& e1 = MNIST& 2.9& 0.12\\
 $TC_5$& a1 = 128& b1 = 0.2& c2 = 200& d1 = Adadelta& e1 = MNIST& 2.5&0.13\\\hline
 \end{tabular}
\end{table*}

\begin{table*}[!htb]
    \centering
        \caption{Testing the ISP on FGSM varying Optimizer (O).}
    \label{tab:t4}
    \begin{tabular}{c|cllll|ll}\hline
    Test Case&  N & R & nb & O & val &loss &accuracy\\\hline
         $TC_1$& 
     a1 = 128& b1 = 0.2& c2 = 10& d1 = Adadelta& e1 = MNIST& 3.08&0.105\\
 $TC_2$& a1 = 128& b1 = 0.2& c2 = 10& d1 = adam& e1 = MNIST& 8.6&0.074\\
 $TC_3$& a1 = 128& b1 = 0.2& c2 = 10& d1 = sgd& e1 = MNIST& 3.85&0.104\\
 $TC_4$& a1 = 128& b1 = 0.2& c2 = 10& d1 = Adagrad& e1 = MNIST&  3.29& 0.092\\
 $TC_5$& a1 = 128& b1 = 0.2& c2 = 10& d1 = RMSProp& e1 = MNIST& 8.48&0.073\\\hline
 \end{tabular}
\end{table*}

Table \ref{tab:t1} varies the number of neurons, indicating how an increase in the model complexity impacts its vulnerability to adversarial examples, with a trend suggesting that more neurons slightly improve the resistance to FGSM attacks, as shown by decreased loss and increased accuracy.

Table \ref{tab:t2} explores different dropout rates, a technique for preventing overfitting. The results demonstrate that both very low and very high dropout rates make the model more susceptible to FGSM, with an optimal range providing better defense.

Table \ref{tab:t3} adjusts the number of classes, testing the model's ability to handle FGSM attacks with varying degrees of classification complexity. However, the impact on model performance does not linearly correlate with the number of classes.

Table \ref{tab:t4} examines the effect of different optimizers on model robustness against FGSM attacks. The choice of optimizer significantly affects the model's defense capability, with some optimizers leading to higher susceptibility.

The emphasis on loss and accuracy metrics in our evaluation framework is pivotal for gauging the effectiveness of adversarial attacks on CNN. These metrics reflect the impact of attacks on model performance, offering a clear picture of how well the network withstands manipulation. As we enhance the model's accuracy through various adjustments, such as optimizing the number of neurons or tweaking the dropout rate, we concurrently observe a improvement in image quality metrics like ERGAS or PSNR. For instance, in Table \ref{tab:t1}, increasing the number of neurons leads to a slight improvement in model accuracy, which correlates with enhancements in image quality metrics, indicating a more robust model against FGSM attacks. This relationship suggests that strategies improving accuracy against adversarial examples also contributes to preserving the integrity of image quality post-attack. Statistical evaluation, using a \textit{paired t-test}, reveals that the improvements in Accuracy, Loss, ERGAS, PSNR, and SSIM are statistically significant ($p < 0.05$), while the improvement in SAM shows promising trends but did not reach statistical significance ($p > 0.05$).

Beyond FGSM, we extended our testing to include other adversarial attack types, DeepFool, PGD, C\&W, and BIM. We observed consistent patterns across these attacks. For example, models with higher dropout rates or those employing certain optimizers like \textit{RMSProp} or \textit{Adam} tended to exhibit more significant performance degradation, as highlighted in Tables \ref{tab:t2} and \ref{tab:t4}. This degradation manifested not only in increased loss and decreased accuracy but also in deteriorated image quality metrics, reinforcing the intertwined relationship between model accuracy and image fidelity in the context of adversarial resilience. The consistent observation of these patterns across different types of attacks validates our approach of using FGSM as a representative example in our evaluation. It demonstrates that the insights gained from FGSM tests offer a reliable indication of how CNN might respond to a broader spectrum of adversarial strategies.

\section{Discussion}
\label{discussion}

Based on the empirical evaluation data, we observe that different adversarial attacks on CNNs demonstrate different impacts when applied to the MNIST or other image type (CIFAR-10, CIFAR-100, and Fashion\_MNIST) datasets. After testing with new hyper-parameter settings such as optimizer = \textit{Adadelta}, $N = 1000$, $R = 0.2$ and $nb = 200$, we improved the accuracy from Table~\ref{tab:fsgm_datasets}'s 0.10 to 0.377, and loss from 3.08 to 1.56 for MNIST. By performing the FGSM attack on a CNN trained on CIFAR-10, CIFAR-100 \& Fashion\_MNIST data sets, we found:
\begin{itemize}
    \item For CIFAR-10, using optimizer \emph{Adadelta}, $esp = 0.01$, and $nb = 500$, improved the accuracy from 0.12 to 0.187, and loss from 6.35 to 2.24. 
    \item For CIFAR-100, using optimizer \emph{sgd}, $esp = 0.01$, and $nb = 200$, improved the accuracy from  0.04 to 0.214, and loss from 6.96 to 3.13. 
    \item For Fashion\_MNIST, using optimizer \emph{Adadelta}, $esp = 0.01$, and $nb = 500$, improved the accuracy from 0.19 to 0.597, and loss from 6.48 to 1.49. 
\end{itemize}

This variety in impact is shown by significant fluctuations in key performance indicators and image quality metrics. These findings underscore the susceptibility of CNNs to sophisticated adversarial methods. It is evident that the development of more robust defense mechanisms is crucial to ensure the reliability and security of CNN applications across different domains.

Table \ref{tab:metric} shows that the DeepFool attack results in the lowest accuracy among all the attacks, indicating a substantial amount of data loss. In contrast, the FGSM attack shows a comparatively better synthesis quality as it records the lowest ERGAS value among the tested attacks. The JSMA attack stands out with the highest peak error. However, it is interesting to note that in terms of SSIM values, which reflect changes in texture, contrast, and luminance, the impacts are quite similar across all attacks on the MNIST dataset. A particularly noteworthy observation is that despite JSMA's high peak-to-peak error, it has the lowest SAM value, suggesting that it maintains higher similarities between the attacked image and the original image. This characteristic of JSMA could be critical for understanding and countering adversarial attacks on CNN models.

The evaluation extends to different test cases for the FGSM attack as detailed in Tables \ref{tab:t1} through \ref{tab:t4}, which illustrate how the accuracy and loss metrics of a CNN are influenced by varying factors such as the number of neurons, the dropout rate, the choice of optimizer, and the number of classes. 


\subsection{Answers to Research Questions}


The answers to our research questions are as follows.
\begin{itemize}
    \item[Q1:] The analysis revealed a variable impact of different adversarial attacks on the classification accuracy of CNNs. Specifically, the DeepFool attack was identified as significantly reducing accuracy due to its effective exploitation of substantial data modifications.
    \item[Q2:] The DeepFool attack stood out as the most effective in inducing the highest error rates, demonstrated by its notably low accuracy scores, highlighting its proficiency in degrading performance metrics.
    \item[Q3:] There was a discernible relationship between image quality metrics and the classification performance of CNNs under attack. For instance, attacks like FGSM, which exhibited lower ERGAS values, suggest a higher quality of image synthesis despite adversarial modifications, indicating an inverse relationship between image quality metrics and classification vulnerability.
    \item[Q4:] Iterative attacks such as BIM and PGD proved more effective than single-step attacks like FGSM. This effectiveness is attributed to the iterative approach's ability to apply perturbations in a refined manner, enhancing the attack's potency.
\end{itemize}

\subsection{Threats to validity}

\textbf{External Validity:} The generalizability of the results is a concern. The study effectively demonstrates the impact of adversarial attacks on CNNs using the MNIST dataset. Moreover, the same results hold forthe datasets like CIFAR-10, CIFAR-100 and Fashion-MNIST. The study provides valuable insights into the vulnerabilities of CNNs, which can inform further research in more varied contexts.

\textbf{Internal Validity:} The causal relationship between the treatment (adversarial attacks) and the observed effects (changes in performance metrics) needs careful examination. Other factors, such as the specific architecture of the CNN or the nature of the dataset, might also influence the outcomes. Ensuring that the observed effects are solely due to the adversarial attacks is crucial for accurate conclusions.

\textbf{Construct Validity:} 
Construct validity is essential in ensuring that the chosen performance metrics, such as accuracy, loss, ERGAS, PSNR, SSIM, and SAM, accurately depict the impact of adversarial attacks on CNNs. The main challenge lies in whether these metrics comprehensively represent the nuanced effects of such attacks, raising concerns about potential misinterpretations that could skew perceptions of CNN vulnerability and resilience. Adopting domain-specific metrics tailored to evaluate adversarial robustness, combined with a multi-dimensional analysis approach that encompasses a broader range of performance indicators, is considered the best option to mitigate these issues. This would offer a more holistic view of a model's behavior under adversarial conditions. Moreover, benchmarking the CNN's performance against baseline models under a variety of attack scenarios, coupled with the use of standardized adversarial robustness testing frameworks, can provide deeper insights into the network's strengths and weaknesses. Ensuring construct validity, therefore, involves a continuous process of validating and updating the assessment methods to align with evolving adversarial techniques, thereby maintaining the accuracy and relevance of conclusions drawn about CNN robustness.

\section{Related Work}
\label{related-work}
Past research in the field of adversarial machine learning has made significant strides. Carlini et al.~\cite{carlini2019evaluating} provide a linearity-based theory for adversarial examples, proposes fast adversarial training, and refutes some alternative hypotheses. The view of adversarial examples as a fundamental property of linear models in high dimensions sparked significant subsequent research into understanding and improving model robustness. Xue et al.~\cite{xue2020machine} provides a comprehensive analysis of contemporary threats to machine learning systems and defenses across the system lifecycle. It highlights open challenges like physical attacks and efficient privacy preservation. The review of evaluations and future directions makes this a wide-ranging resource for security in machine learning.

Xu et al.~\cite{xu2020adversarial} provide a comprehensive review of adversarial attacks and defenses across multiple modalities including images, graphs, and text. 
The analysis of various attacks and security testing methods provides a foundation for choosing CNN hyperparameters that enhance resilience to adversarial attacks in image processing. Goodfellow et al.~\cite{goodfellow2014explaining} outline a set of principles for evaluating the robustness of machine learning defenses against adversarial examples, emphasizing the importance of a well-defined threat model and skepticism towards one's own results. It advocates for rigorous testing using adaptive attacks, caution against security through obscurity, and the necessity of public code and model release for reproducibility. Additionally, they provide a checklist to avoid common pitfalls in such evaluations, encouraging comprehensive testing and comparison with existing work.

Wu and Zhu~\cite{wu2020towards} provide useful insights into factors influencing adversarial transferability and proposes a simple but effective smoothed gradient attack to enhance it. The attack has implications on evaluating model robustness. Recent advancements, such as a cluster-based approach with a dynamic reputation system for Flying Ad hoc Networks~\cite{gupta2024scfs} and a weighted, spider monkey-based optimization method for Vehicular Ad hoc Networks~\cite{gupta2024novel}, have shown significant improvements in performance metrics, security, and reliability for CNN.

Our work aims to expand upon these foundations by specifically focusing on the impact of white-box adversarial attacks on CNN performance metrics. Unlike previous studies that broadly addressed adversarial threats in machine learning, our research delves into the detailed analysis of how these attacks affect CNNs, providing a more focused understanding of their vulnerabilities and potential defenses. This specificity in studying the direct effects of attacks on CNNs sets our work apart and underscores its importance in the broader context of machine learning security.

\section{Conclusions and Future work}
\label{conclusion}


We showed that Convolutional Neural Networks exhibit significant vulnerability to a range of adversarial attacks, which lead to notable degradation in performance metrics like accuracy, loss, and image quality. The research underscores the importance of developing more resilient CNN architectures and defense mechanisms to counteract these vulnerabilities, particularly in critical applications where CNN reliability is paramount. The findings provide valuable insights for future research aimed at enhancing the security and robustness of CNNs against sophisticated adversarial threats.

These test scenarios plays a crucial role in the development and refinement of defense mechanisms for Convolutional Neural Networks (CNNs). By subjecting models to a wide range of attack scenarios, we can observe the specific ways in which adversarial inputs manipulate model behavior. This insight is invaluable for devising defense strategies that directly counteract the observed vulnerabilities. For instance, testing can reveal if a model is particularly sensitive to slight perturbations in certain input features, leading to the development of input preprocessing or feature squeezing techniques as countermeasures. Similarly, the effectiveness of adversarial training can be assessed and optimized through iterative testing, by incorporating a diverse set of adversarial examples generated from the latest attack methods. Moreover, testing helps in evaluating the practicality of defense mechanisms under real-world conditions, ensuring that they do not unduly compromise model accuracy or performance. Through this iterative process of attack simulation, vulnerability assessment, and defense implementation, testing fosters a deeper understanding of adversarial threats and guides the creation of more robust and resilient systems.


\section*{Acknowledgment}

This project was supported in part by the US National Science Foundation under award number OAC 1931363, and Grant No. 1822118 and 2226232.



%

\bibliographystyle{IEEEtran}
\bibliography{main}

\end{document}